\documentclass[letterpaper,english,preprint,aps,nofootinbib,superscriptaddress]{revtex4-1}
\usepackage[T1]{fontenc}
\usepackage[utf8]{inputenc}
\usepackage{amsmath}
\usepackage{amssymb}

\usepackage{color}
\usepackage{graphicx,subfig}

\makeatletter

\pdfpageheight\paperheight
\pdfpagewidth\paperwidth

\providecommand{\LyX}{L\kern-.1667em\lower.25em\hbox{Y}\kern-.125emX\@}

\makeatletter

\pdfpageheight\paperheight
\pdfpagewidth\paperwidth

\providecommand{\LyX}{L\kern-.1667em\lower.25em\hbox{Y}\kern-.125emX\@}

\makeatother

\usepackage{babel}
\begin{document}

\title{The holographic p+ip solution failed to win the competition\\
in dRGT massive gravity
} \vskip 2cm \vskip 2cm

\author{Zhang-Yu Nie}
\email{niezy@kust.edu.cn}
\affiliation{School of Science, Kunming University of Science and Technology, Kunming 650500, China}

\author{Ya-Peng Hu}
\email{huyp@nuaa.edu.cn}
\affiliation{College of Science, Nanjing University of Aeronautics and Astronautics, Nanjing 210016, China}

\author{Hui Zeng}
\email{zenghui@kust.edu.cn}
\affiliation{School of Science, Kunming University of Science and Technology, Kunming 650500, China}

\date{\today}
\begin{abstract}
In this paper, the holographic p-wave superfluid model with charged complex vector field is studied in dRGT massive gravity beyond the probe limit. The stability of p-wave and p+ip solutions are compared in the grand canonical ensemble. The p-wave solution always get lower value of grand potential than the p+ip solution, showing that the holographic system still favors an anisotropic (p-wave) solution even with considering a massive gravity theory in bulk. In the holographic superconductor models with dRGT massive gravity in bulk, a key sailing symmetry is found to be violated by fixing the reference metric parameter $c_0$. Therefore, in order to get the dependence of condensate and grand potential on temperature, different values of horizon radius should be considered in numerical work. With a special choice of model parameters, we further study the dependence of critical back-reaction strength on the graviton mass parameter, beyond which the superfluid phase transition become first order. We also give the dependence of critical temperature on the back reaction strength $b$ and graviton mass parameter $m^2$.
\end{abstract}



\maketitle
\flushbottom

\section{\bf Introduction}
The AdS/CFT correspondence~\cite{Maldacena:1997re,Gubser:1998bc,Witten:1998qj} provides a novel way to study the strongly coupled systems. One successful application is the so called holographic superconductor~\cite{Gubser:2008px,Hartnoll:2008vx} , which mimic the superconductor phase transition with a spontaneously emerged charged hair in the bulk black hole spacetime. Various different matter fields as well as gravitational theories are considered to build different holographic superconductor and superfluid models~\cite{Cai:2015cya}, in order to realize various superconducting phenomenons and to check some universal laws~\cite{Zaanen:2015oix}.

Superfluid with p-wave pairing has also been realized holographically. In the early study~\cite{Gubser:2008wv}, an SU(2) gauge field is introduced to realize both the electro-magnetic field and the condensed vector order. Recent studies~\cite{Cai:2013aca,Cai:2013pda,Li:2013rhw,Wu:2014bba,Pan:2015lit,Lai:2016yma,Lu:2018tdo,Lu:2020phn,Huang:2019yov,Lv:2020ecm} reconsidered the holographic p-wave model with a charged massive vector(also called Proca field) in bulk and get more interesting phase transition phenomenon.

Lessons on superfluid Helium-3 tell us that the superfluid phases with p-wave pairing exhibit various spatial symmetry~\cite{Volhardt-Wolfle-1990}. Therefore it is interesting to study possible new phases in holographic p-wave model. Previous study~\cite{Nie:2016pjt} already imply that a kind of p+ip solution can be realized. In probe limit, the p-wave and p+ip solution get the same value of grand potential and thus form a degenerate state. However, when back-reaction on the metric is turned on, the p+ip solution always get a larger value of grand potential than the p-wave solution which is more stable. In order to study the properties such as conductivity and chiral magnetic effect in the p+ip phase, it would be necessary to get stable p+ip solution at first. A possible approach is extending this study in more general theories of gravity. This can also help us to understand whether the spacetime favors isotropy(p+ip solution) or anisotropy(p-wave solution) in various theories of gravity.

Recently, a ghost free gravity theory with massive graviton is proposed in Ref.~\cite{deRham:2010kj}. The holographic dual of this massive gravity theory show translational symmetry breaking effects~\cite{Vegh:2013sk}. The holographic superconductor model with s-wave paring also has been studied in this massive gravity theory~\cite{Zeng:2014uoa}, giving a finite value of conductivity at zero frequency. Since the massive gravity theory has non-trivial effects in the holographic study, it would be interesting to study the problem of competition between p-wave and p+ip orders. Some other interesting studies on dRGT massive gravity can be found in Refs.~\cite{Cai:2014znn,Hu:2016mym,Hu:2015dnl,Hu:2015xva,Xu:2015rfa,Hu:2017nzw}.

In this paper, we study the p-wave and p+ip solutions in a holographic model with charged complex vector field in dRGT massive gravity. We work in the grand canonical ensemble and compare the grand potential of the two solutions with considering the back-reaction of matter fields on metric. We also show the effect of graviton mass on this system. The rest of the paper is organized as follows. In Sec.~\ref{sect:setup} we give the set up of the new p-wave model in the massive gravity. In Sec.~\ref{sect:results} we show the results of the stability problem between p-wave and p+ip solutions as well as the effect of graviton mass parameter. Finally, we conclude the main results in this paper and give some discussions in Sec.~\ref{sect:conclusion}.
\section{The holographic p-wave model from massive gravity}\label{sect:setup}
In this section, we give details of the setup of holographic p-wave model with charged complex vector field in massive gravity. We also give the expression for condensate as well as grand potential of the p-wave and p+ip solutions.
\subsection{The model setup}
The action can be expressed as
\begin{eqnarray}
S  &=&S_G+S_M, \label{Sall}\\
S_G&=&\frac{1}{2 \kappa_g ^2}\int d^{4}x \sqrt{-g} \big(R-2\Lambda +m^2 \sum_i^4 c_i \mathcal{U}_i \big),  \label{Sgravity}\\
S_M&=&\frac{1}{q^2}\int d^{4}x \sqrt{-g}\Big(-\frac{1}{4}F_{\mu\nu}F^{\mu\nu}-\frac{1}{2} \rho^\dagger_{\mu\nu}\rho^{\mu\nu}-m_p^2 \rho^\dagger_\mu \rho^\mu \Big). \label{Smatter}
\end{eqnarray}
The total action of this system can be divided into the gravity part and the matter part. Eq.~\eqref{Sgravity} is the expression for gravity part, in which the last term is the mass term for graviton. The $c_i$ are constants and $\mathcal{U}_i$ are symmetric polynomials of the eigenvalues of the $4\times 4$ matrix $\mathcal{K} ^\mu_\nu=\sqrt{g^{\mu \alpha}f_{\alpha \nu}}$
\begin{eqnarray}
\mathcal{U}_1&=&[\mathcal{K}],					\nonumber \\
\mathcal{U}_2&=&[\mathcal{K}]^2-[\mathcal{K}^2],		\nonumber \\
\mathcal{U}_3&=&[\mathcal{K}]^3-3[\mathcal{K}][\mathcal{K}^2]+2[\mathcal{K}^3],		\nonumber \\
\mathcal{U}_4&=&[\mathcal{K}]^4-6[\mathcal{K}^2][\mathcal{K}]^2+8[\mathcal{K}^3][\mathcal{K}]+3[\mathcal{K}^2]^2-6[\mathcal{K}^4].
\end{eqnarray}
The square brackets denote the trace $[\mathcal{K}]=\mathcal{K}_\mu^\mu$.

The action of the matter part~\eqref{Smatter} includes the U(1) gauge field $A_\mu$ as well as the massive complex vector field $\rho_\mu$ charged under $A_\mu$~\cite{Cai:2013aca,Cai:2013pda}. The field strength of these two fields are $F_{\mu\nu}=\nabla_\mu A_\nu-\nabla_\nu A_\mu$ and $\rho_{\mu\nu}=D_\mu \rho_\nu-D_\nu \rho_\mu$ respectively, where $D_\mu =\nabla_\mu -i A_\mu$. The superscript ``$^\dagger$" means complex conjugate, and $m_p$ is the mass for the vector field and controls the conformal dimension of the p-wave order.

The equations of motion for this coupled system can be expressed as those for the matter fields
\begin{eqnarray}
\nabla^\nu F_{\nu\mu} &= &i (\rho^\nu\rho_{\nu\mu}^\dagger-\rho^{\nu\dagger}\rho_{\nu\mu}),\\
D^\nu \rho_{\nu\mu}  - m_p^2 \rho_\mu &=& 0,
\end{eqnarray}
and the Einstein equations for the metric
\begin{equation}
R_{\mu\nu} -\frac{1}{2}(R-2\Lambda) g_{\mu\nu} +m^2\mathcal{X}_{\mu\nu}= b^2 \mathcal{T}_{\mu\nu},
\end{equation}
where $b= \kappa_g/q$ characterizes the strength of back reaction of the matter fields on the background geometry and the tensor $\mathcal{X}_{\mu\nu}$ is
\begin{eqnarray}
\mathcal{X}_{\mu\nu}=&&-\frac{c_1}{2}(\mathcal{U}_1 g_{\mu\nu}-\mathcal{K}_{\mu\nu})
-\frac{c_2}{2}(\mathcal{U}_2 g_{\mu\nu}-2\mathcal{U}_1 \mathcal{K}_{\mu\nu} +2 \mathcal{K}_{\mu\nu}^2) 	\nonumber \\ &&
-\frac{c_3}{2}( \mathcal{U}_3 g_{\mu\nu}-3\mathcal{U}_2 \mathcal{K}_{\mu\nu}+6\mathcal{U}_1\mathcal{K}_{\mu\nu}^2-6\mathcal{K}_{\mu\nu}^3)	\nonumber \\ &&
-\frac{c_4}{2}(\mathcal{U}_4 g_{\mu\nu}-4\mathcal{U}_3 \mathcal{K}_{\mu\nu}+12\mathcal{U}_2\mathcal{K}_{\mu\nu}^2-24\mathcal{U}_1 \mathcal{K}_{\mu\nu}^3+24\mathcal{K}_{\mu\nu}^4).
\end{eqnarray}
$\mathcal{T}_{\mu\nu}$ is the stress-energy tensor of the matter sector
\begin{eqnarray}
\mathcal{T}_{\mu\nu} = &(-\frac{1}{4}F_{\mu\nu}^a F^{a\mu\nu} -\frac{1}{2} \rho^\dagger_{\mu\nu|}\rho^{\mu\nu}-m_p^2 \rho^\dagger_\mu \Psi^\mu) g_{\mu\nu} +F_{\mu\lambda} F_\nu^\lambda  \nonumber\\
& + \rho^\dagger_{\mu\lambda}\rho_\nu^\lambda+\rho^\dagger_{\nu\lambda}\rho_\mu^\lambda+m_p^2 (\rho^\dagger_\mu \rho_\nu+ \rho^\dagger_\nu \rho_\mu).
\end{eqnarray}

If the matter fields are turned off, the massive gravity action admits analytical solutions in the form~\cite{Vegh:2013sk}
\begin{eqnarray}
&ds^2&=g_{\mu\nu}dx^\mu dx^\nu=-N(r)dt^2+\frac{1}{N(r)}dr^2+r^2h_{ij}dx^idx^j,\\
&&=-N(r)dt^2+\frac{1}{N(r)}dr^2+\frac{r^2}{L^2}(dx^{2}+ dy^2),\\
&N(r)&=c_0^2 c_2 m^2+\frac{1}{2} c_0 c_1 m^2 r-\frac{2 M_0}{L^2 r}+\frac{r^2}{L^2} ,
\end{eqnarray}
while the reference metric $f_{\mu\nu}$ is taken as
\begin{eqnarray}\label{ReferenceM}
f_{\mu\nu}=\text{diag}(0,0,c_0^2 h_{ij}).
\end{eqnarray}

We wish to study the p-wave and p+ip solutions dual to superfluid phases where the U(1) symmetry is spontaneously broken. Therefore we take the ansatz for matter fields as
\begin{eqnarray}
A_t=\phi(r),~\rho_x=\Psi_x(r),~\rho_y=i \Psi_y(r). \label{MatterAnsatz}
\end{eqnarray}
A metric ansatz~\cite{Nie:2014qma,Ammon:2009xh} consistent with this matter ansatz can be given as
\begin{eqnarray}\label{fullmetric}
&ds^2=-N(r) \sigma (r)^2dt^2+\frac{1}{N(r)}dr^2+\frac{r^2}{L^2}(\frac{1}{f(r)^2}dx^{2}+f(r)^2 dy^2),
\end{eqnarray}
with
\begin{equation}
N(r)=c_0^2 c_2 m^2+\frac{1}{2} c_0 c_1 m^2 r-\frac{2 M(r)}{L^2 r}+\frac{r^2}{L^2} . 
\end{equation}
We still take the same style of reference metric $f_{\mu\nu}$~(\ref{ReferenceM}), where the expression for $h_{ij}$ change to be
\begin{eqnarray}
h_{ij}dx^idx^j=\frac{1}{L^2}(\frac{1}{f(r)^2}dx^{2}+f(r)^2 dy^2).
\end{eqnarray}

With the above matter and metric ansatz, we can get the full equations of motion as
\begin{eqnarray}\label{EoMs1}
M'(r)&=& \nonumber
\frac{b^2 L^4 \phi (r)^2}{2 N(r) \sigma (r)^2}  \Big(f(r)^2 \Psi_x(r)^2+\frac{\Psi_y(r)^2}{f(r)^2} \Big)
+\frac{1}{2} b^2 L^4  N(r)   \Big(f(r)^2\Psi_x'(r)^2 +\frac{\Psi_y'(r)^2}{f(r)^2} \Big) \\&& \nonumber
+\frac{1}{2}b^2 L^4 m_p^2  \Big(f(r)^2 \Psi_x(r)^2+\frac{\Psi_y(r)^2}{f(r)^2}\Big)  \\&&
+\frac{b^2 L^2 r^2 \phi '(r)^2}{4 \sigma (r)^2} +\frac{L^2 r^2 N(r) f'(r)^2}{2 f(r)^2},
\\
\label{EoMs2} \sigma '(r)&=&\frac{b^2 L^2 \phi (r)^2}{r N(r)^2 \sigma (r)}\Big(f(r)^2 \Psi_x(r)^2+\frac{\Psi_y(r)^2}{f(r)^2} \Big)
+\frac{b^2 L^2 \sigma (r) }{r}\Big(f(r)^2\Psi_x'(r)^2 +\frac{\Psi_y'(r)^2}{f(r)^2} \Big)  \nonumber\\&&
+\frac{r \sigma (r) f'(r)^2}{f(r)^2},
\\
\label{EoMs3} f''(r)&=&-\frac{b^2 L^2 f(r) \phi (r)^2}{r^2 N(r)^2 \sigma (r)^2} \Big(f(r)^2 \Psi_x(r)^2-\frac{\Psi_y(r)^2}{f(r)^2} \Big)
+\frac{b^2 L^2 f(r)}{r^2}\Big(f(r)^2\Psi_x'(r)^2 -\frac{\Psi_y'(r)^2}{f(r)^2} \Big)\nonumber\\&&
+\frac{b^2 L^2 m_p^2 f(r)}{r^2 N(r)}\Big(f(r)^2 \Psi_x(r)^2-\frac{\Psi_y(r)^2}{f(r)^2} \Big)\nonumber\\&&
+\frac{f'(r)^2}{f(r)}
-\frac{f'(r) N'(r)}{N(r)}-\frac{f'(r) \sigma '(r)}{\sigma (r)}-\frac{2 f'(r)}{r},
\\
\label{EoMs4} \phi ''(r)&=&\Big(\frac{\sigma '(r)}{\sigma (r)}-\frac{2 }{r} \Big)\phi '(r)
+\frac{2 L^2}{r^2 N(r)}\Big(f(r)^2 \Psi_x(r)^2+\frac{\Psi_y(r)^2}{f(r)^2} \Big)  \phi (r),\\
\label{EoMs5} \Psi_x''(r)&=&-\Big( \frac{N'(r)}{N(r)}+\frac{\sigma'(r)}{\sigma(r)}+\frac{2 f'(r)}{f(r)} \Big) \Psi_x'(r)
-\Big( \frac{\phi(r)^2}{N(r)^2 \sigma(r)^2}-\frac{m_p^2}{N(r)} \Big) \Psi_x(r),\\
\label{EoMs6} \Psi_y''(r)&=&-\Big( \frac{N'(r)}{N(r)}+\frac{\sigma'(r)}{\sigma(r)}-\frac{2 f'(r)}{f(r)} \Big) \Psi_y'(r)
-\Big( \frac{\phi(r)^2}{N(r)^2 \sigma(r)^2}-\frac{m_p^2}{N(r)} \Big) \Psi_y(r).
\end{eqnarray}

We also need to specify boundary conditions in order to solve this set of equations numerically, both on the horizon and on the $r\rightarrow \infty$ boundary of bulk AdS black brane spacetime. The boundary behaviors near horizon can be expressed as
\begin{eqnarray}\label{BC-horizon}
M(r)		&=&	\frac{1}{2}+M_{h1}(r-r_h)+...		\\
\sigma(r)	&=&	\sigma_{h0}+\sigma_{h1}(r-r_h)+...	\\
f(r)		&=&	f_{h0}+f_{h1}(r-r_h)+...	\\
\phi(r)	&=&	\phi_{h1}(r-r_h)+\phi_{h2}(r-r_h)^2+...	\\
\Psi_x(r)	&=&	\Psi_{xh0}+\Psi_{xh1}(r-r_h)+...	\\
\Psi_y(r)	&=&	\Psi_{yh0}+\Psi_{yh1}(r-r_h)+...
\end{eqnarray}
Where the independent parameters are $(\sigma_{h0},~f_{h0},~\phi_{h1},~\Psi_{xh0},~\Psi_{yh0})$.

The expansions near the $r\rightarrow \infty$ boundary are
\begin{eqnarray}\label{BC-boundary}
M(r)		&=&	M_{b0}+\frac{M_{b1}}{r}+...		\\
\sigma(r)	&=&	\sigma_{b0}+\frac{\sigma_{b1}}{r}+...	\\
f(r)		&=&	f_{b0}+\frac{f_{h1}}{r}+...	\\
\phi(r)	&=&	\mu-\frac{\rho}{r}+...	\\
\Psi_x(r)	&=&	\frac{\Psi_{x-}}{r^{\Delta_-}}+\frac{\Psi_{x+}}{r^{\Delta_+}}+...	\\
\Psi_y(r)	&=&	\frac{\Psi_{y-}}{r^{\Delta_-}}+\frac{\Psi_{y+}}{r^{\Delta_+}}+...~~,
\end{eqnarray}
where
\begin{equation}
\Delta_\pm=(1\pm\sqrt{1+4m_p^2 L^2})/2
\end{equation}
are the conformal dimensions of the source and expectation value of the dual vector operator. In this study, some constraints on the boundary coefficients $(\Psi_{x-}=0,~\Psi_{y-}=0,~\sigma_{b0}=1,~f_{b0}=1)$ are introduced to confirm the solutions to be asymptotically AdS and dual to a source free condensed phase.

The above knowledge tell us that with the constraints from boundary, we can get a set of solutions characterized by one parameter $\mu$. There are also parameters including $(L,~b,~m,~m_p,~c_0,~c_1,~c_2)$ which should be fixed before the numerical work.

There are several scaling symmetries in this model
\begin{eqnarray}
(1)~& \Psi_x \rightarrow \lambda^2 \Psi_x,~\Psi_y \rightarrow \lambda^2 \Psi_y,~\phi \rightarrow \lambda^2 \phi,~ N \rightarrow \lambda^2 N,~ \nonumber \\
& m_p \rightarrow \lambda m_p,~ L \rightarrow \lambda^{-1} L,~ b \rightarrow \lambda^{-1} b,~m \rightarrow \lambda m;\\
(2)~& \Psi_x \rightarrow \lambda \Psi_x,~\Psi_y \rightarrow \lambda \Psi_y,~\phi \rightarrow \lambda \phi,~ N \rightarrow \lambda^2 N,~ \nonumber \\
& M \rightarrow \lambda^3 M,~ r \rightarrow \lambda r,~ c_0 \rightarrow \lambda c_0; \label{scaling2}\\
(3)~&\phi \rightarrow \lambda \phi,~ \sigma \rightarrow \lambda \sigma;\\
(4)~& \Psi_x \rightarrow \lambda^{-1} \Psi_x,~\Psi_y \rightarrow \lambda \Psi_y,~ f \rightarrow \lambda f;\\
(5)~& c_0 \rightarrow \lambda c_0,~c_1 \rightarrow \lambda^{-1} c_1,~ c_2 \rightarrow \lambda^{-2} c_2;\\
(6)~& c_0 \rightarrow \lambda c_0,~c_1 \rightarrow \lambda c_1,~ m \rightarrow \lambda^{-1} m.
\end{eqnarray}

The first four symmetries are similar to those scaling symmetries in previous study in system without graviton mass term, and the last two only involve the parameters in graviton mass term. It is the second scaling symmetry Eq.~(\ref{scaling2}) that usually be used to get the varying values of temperature of the condensed solutions after solving these equations with a fixed value of horizon radius. However, in the massive gravity case this scaling symmetry involves the parameter $c_0$, therefore the value of $c_0$ changes with temperature if we use the same trick. To get solutions with the fixed value of $c_0$ and varying temperature, we apply a different  numerical treatment in which we fix the chemical potential $\mu$ while the horizon radius $r_h$ can be tuned to get different values of temperature with fixed value of $c_0$. This numerical treatment is more general than using the usual trick.
\subsection{Condensates of p-wave and p+ip soluitions}
With the standard shooting method and our new numerical treatment, we can get solutions dual to the ordinary p-wave states and the p+ip one respectively. In the p-wave solution we have $(\Psi_x=\Psi_p(r),~\Psi_{y}(r)=0$(or equivalently $\Psi_x=0,~\Psi_{y}(r)=\Psi_p(r)$), and in the p+ip solution we have $\Psi_x(r)=\Psi_y(r)=\Psi_{pip}(r)$. According to the AdS/CFT dictionary, the condensed value of the orders are equal to $\Psi_{x+}$ and $\Psi_{y+}$ respectively. In order to better comparing the condensed value of the two different solutions, an expression of condensed value applicable for both the two solutions is~\cite{Nie:2016pjt}
\begin{eqnarray}
\Psi_+=\sqrt{\Psi_{x+}^2+\Psi_{y+}^2}~.
\end{eqnarray}
One can calculate the energy momentum tensor of the matter fields to confirm that the p-wave solution is anisotropic while the p+ip solution is isotropic (in AdS$^4$)~\cite{Nie:2016pjt}. In this sense, the stability relation between the two solutions also give some insights of the favor of the gravitational theory between isotropy and anisotropy.

The temperature of the boundary system is duel to the Hawking temperature of the bulk black brane
\begin{eqnarray}
T&=&\frac{N'(r_h)\sigma (r_h)}{4 \pi}   \nonumber\\
&=&\frac{3 \sigma_{h0} r_h}{4 \pi}-\frac{b^2 \phi_{h1}^2}{8 \pi \sigma_{h0} r_h}+\frac{c_0 c_1 m^2\sigma_{h0}}{4 \pi}+\frac{c_0^2 c_2 m^2\sigma_{h0}}{4 \pi r_h} \nonumber \\
&&-\frac{b^2 m_p^2 \sigma_{h0}}{4 \pi r_h}(f_{h0}^2 \Psi_{xh0}^2+\frac{\Psi_{yh0}^2}{f_{h0}^2})~. \label{Temperature}
\end{eqnarray}

As we have explained in the previous subsection, the second scaling symmetry Eq.~(\ref{scaling2}) involve the parameter $c_0$ and we can not easily get the condensed solutions with varying temperature by using the scaling trick. In order to solve this problem, we explore new numerical technic to get solutions with a varying horizon radius $r_h$ and fixed value of chemical potential $\mu =3$.

We can draw condensed value of the both solution with respect to temperature $T$ once we fixed a set of values $(L,~b,~m,~m_p,~c_0,~c_1,~c_2)$ and solved the equations of motion numerically. In this work we focus on the effect of graviton mass, we set $L=1$ and $c_0=c_1=-2 c_2=1$ for simplicity~\cite{Zeng:2014uoa,Hu:2015dnl,Hu:2016mym}.

If we take probe limit $b\rightarrow 0$, because the symmetry between the equations of motion for $\Psi_x$ and $\Psi_y$, the p-wave and p+ip solutions will have the same value of critical temperature, condensate as well as grand potential~\cite{Nie:2016pjt}. If we go beyond the probe limit with finite value of $b$, the degenerate p-wave and p+ip solutions will still have the same values of critical temperature, but the condensate and grand potential curves of the two solutions will be split gradually away from the critical point. In Einstein gravity, the p-wave solution always have a lower grand potential, the p+ip solution only could be stable when the p-wave solution does not exist in that region.
\subsection{Grand potential of p-wave and p+ip soluitions}
We wish to study the stability problem between the p wave and p+ip solutions in this massive gravity setup, and it is necessary to calculate the grand potential of this system. We work in the grand canonical ensemble and the grand potential is given by the Euclidean on-shell action.
\begin{eqnarray}\label{Omega-SE}
\Omega=T S_E~.
\end{eqnarray}
Besides the bulk action~(\ref{Sall}), a Gibbons-Hawking term
\begin{eqnarray}\label{S-GH}
S_{GH}&=&-\frac{1}{\kappa_g ^2}\int_{\Sigma } d^{3}x \sqrt{-\gamma} K ~.
\end{eqnarray}
as well as a counter term 
\begin{eqnarray}\label{S-ct}
S_{ct}&=&-\frac{1}{\kappa_g ^2}\int_{\Sigma } d^{3}x \sqrt{-\gamma} \Big( \frac{2}{L}+\frac{1}{2}R[\gamma]
+\frac{1}{4}m^2 L \big(c_1\mathcal{U}_1-\frac{1}{16}L^2m^2c_1^2 \mathcal{U}_1^2+2c_2\mathcal{U}_2 \big)       \Big), ~~~~~
\end{eqnarray}
where $\Sigma$ denotes the boundary hyper surface at $r \rightarrow \infty$ and $R[\gamma]$ is the Ricci scalar of the induced metric $\gamma_{\mu\nu}$ on $\Sigma$, should also be included~\cite{Cao:2015cza}.

With our metric and matter ansatz~(\ref{MatterAnsatz},~\ref{fullmetric}), the expression for the grand potential density $\Omega$ is
\begin{eqnarray}\label{Omega}
\kappa_g^2 V_2 \Omega &=& \int_{r_h}^\infty \big( \frac{c_0^2 c_2 m^2 \sigma(r)}{L^2}+ \frac{c_0 c_1 m^2 r \sigma(r)}{2 L^2} \big) dr - \frac{r N(r) \sigma(r)}{L^2}\Big|_{r=\infty} \nonumber \\
&&+\sqrt{N(r)}  \frac{\big(r^2 \sqrt{N(r)}\sigma(r) \big)'}{L^2} \Big|_{r=\infty} \nonumber \\
&&+\sqrt{N(r)} \sigma(r) \Big(\frac{c_0 c_1 m^2 r- 2 c_0^2 c_2 m^2}{2L}+ \frac{c_0^2 c_1^2 m^4 L}{16} -\frac{2r^2}{L^3}\Big) \Big|_{r=\infty}~,
\end{eqnarray}
where $V_2=\int dxdy$ is the volume of the boundary system.

The terms in the first line are the contribution from bulk action, the term in the second line is from the Gibbons-Hawking term and the last line show the contribution from boundary counter terms. Both the bulk integration term and the boundary terms are divergent, but the sum of the two is convergent. We take the boundary term into the integration to get a convergent result in our numerical work
\begin{eqnarray}\label{OmegaIntegral}
\kappa_g^2 V_2 \Omega &=& \int_{r_h}^\infty \big( \frac{c_0^2 c_2 m^2 \sigma(r)}{L^2}+ \frac{c_0 c_1 m^2 r \sigma(r)}{2 L^2} \big) dr  \nonumber \\
&&+\int_{r_h}^\infty\Big[ - \frac{r N(r) \sigma(r)}{L^2}+\sqrt{N(r)}  \frac{\big(r^2 \sqrt{N(r)}\sigma(r) \big)'}{L^2}  \nonumber \\
&&+\sqrt{N(r)} \sigma(r) \Big(\frac{c_0 c_1 m^2 r- 2 c_0^2 c_2 m^2}{2L}+ \frac{c_0^2 c_1^2 m^4 L}{16} -\frac{2r^2}{L^3}\Big) \Big]' dr \nonumber \\
&&+\Big[ - \frac{r N(r) \sigma(r)}{L^2}+\sqrt{N(r)}  \frac{\big(r^2 \sqrt{N(r)}\sigma(r) \big)'}{L^2}  \nonumber \\
&&+\sqrt{N(r)} \sigma(r) \Big(\frac{c_0 c_1 m^2 r- 2 c_0^2 c_2 m^2}{2L}+ \frac{c_0^2 c_1^2 m^4 L}{16} -\frac{2r^2}{L^3}\Big) \Big]\Big|_{r=r_h} ~.
\end{eqnarray}
The final expression is 
\begin{eqnarray}\label{FinalOmega}
\kappa_g^2 V_2 \Omega
&=&  \int_{r_h}^\infty \Big( \frac{N(r)\sigma(r)}{L^2}-\frac{4 r \sqrt{N(r)}\sigma(r)}{L^3}  +\frac{2r\sigma(r)N'(r)}{L^2}+\frac{3r N(r)\sigma'(r)}{L^2}   \nonumber\\
&&+\frac{r^2\sigma(r)N'(r)}{L^3\sqrt{N(r)}} +\frac{2r^2\sqrt{N(r)}\sigma'(r)}{L^3} +\frac{2r^2N'(r)\sigma'(r)}{2L^2}+\frac{r^2\sigma(r)N''(r)}{2L^2}+\frac{r^2N(r)\sigma''(r)}{L^2}  \nonumber\\
&&\frac{c_0^2 c_2 m^2 \sigma(r)}{L^2}+ \frac{c_0 c_1 m^2 r \sigma(r)}{2 L^2} - \frac{c_0 c_1 m^2 \sqrt{N(r)} \sigma(r)}{2 L} -\frac{c_0 c_1 m^2 r \sqrt{N(r)}\sigma'(r)}{2 L} \nonumber\\
&&-\frac{c_0 c_1 m^2 r \sigma(r) N'(r)}{4L\sqrt{N(r)}} -\frac{c_0^2 c_2 m^2\sqrt{N(r)}\sigma'(r)}{L} -\frac{c_0^2 c_2 m^2\sigma(r) N'(r)}{2 L \sqrt{N(r)}}  \nonumber\\
&& +\frac{c_0^2 c_1^2 m^2 L \sqrt{N(r)}\sigma'(r)}{16} +\frac{c_0^2 c_1^2 m^2 L \sigma(r) N'(r)}{32 \sqrt{N(r)}}    \Big) dr   +\frac{r^2\sigma(r)N'(r)}{2L^2} \Big|_{r=r_h} ~.
\end{eqnarray}

With the above formulas in hand, we studied the competition between the p-wave and p+ip solutions as well as the phase structure of this holographic system. We show our main results in the next section.

\section{Competition between the two solutions and the influence of $m^2$ on $T_c$}\label{sect:results}
\begin{figure}\center
\includegraphics[height=4cm] {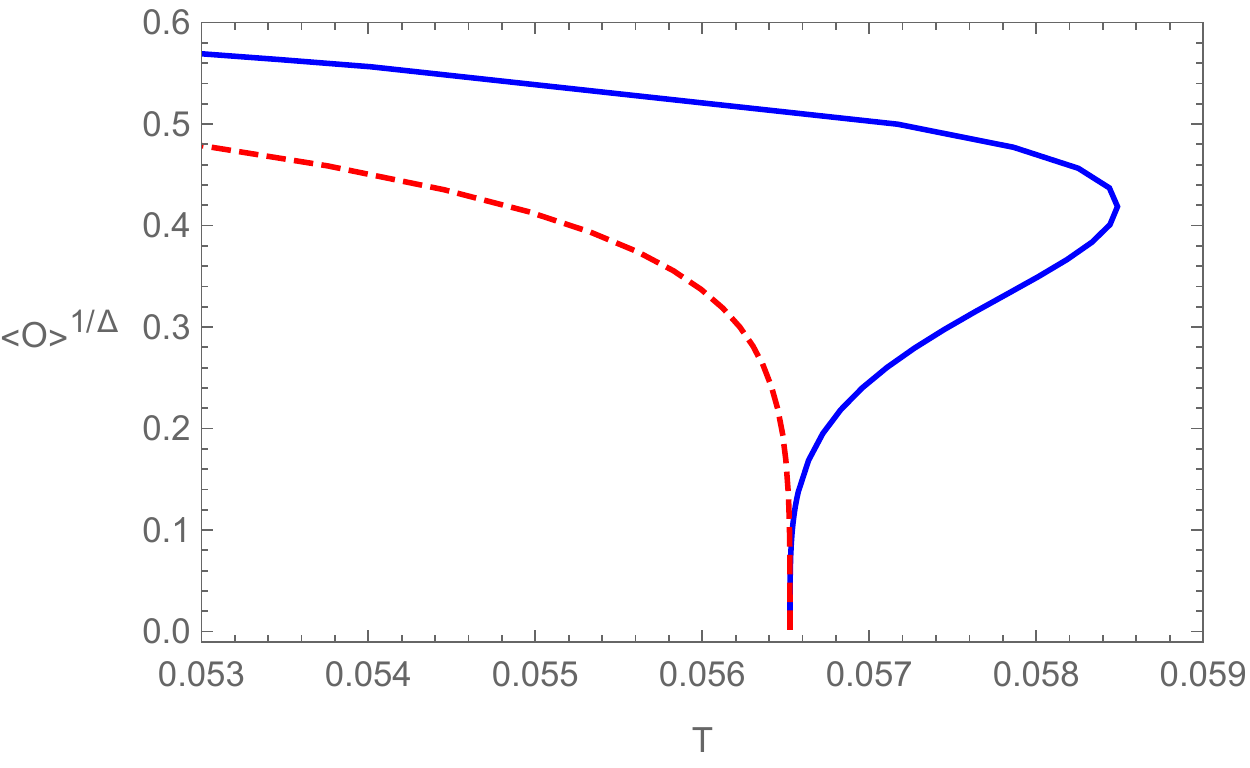}  \quad \quad
\includegraphics[height=4cm] {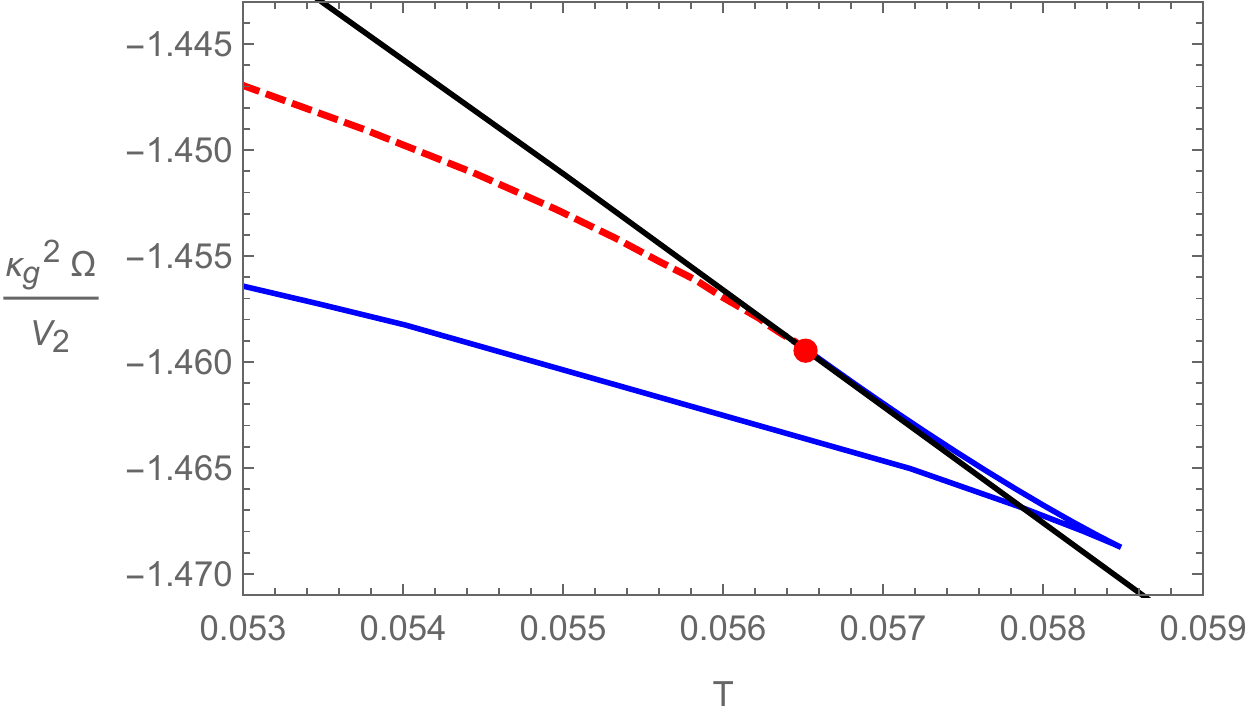}
\caption{\label{CondensateFreeE} Then condensate(Left Plot) and grand potential(Right Plot) curves for the p-wave and p+ip solutions with $m^2=0.3$ and $b=0.68$. We use solid blue line to denote the curves for p-wave solution and the dashed red line for the p+ip solution. The solid black line denote the grand potential curve for the normal solution without any condensate.
}
\end{figure}

We get the two solutions numerically and compared the grand potential of the two. Unfortunately, the p+ip solution still failed to win the competition with the choice of parameters we considered. The figures of temperature dependence of condensate value as well as grand potential are qualitatively the same to the results in Einstein gravity~\cite{Nie:2016pjt}. Therefore we only show a typical case, in which the p-wave solution is a first order phase transition while the p+ip one is still second order, with $m^2=0.3$ and $b=0.68$ in Figure.~\ref{CondensateFreeE}.

We can see from Figure.~\ref{CondensateFreeE} that the p-wave and p+ip solutions share the same critical point, which can be explained by the degeneration of the two solutions in probe limit~\cite{Nie:2016pjt}. The p-wave solution with a first order phase transition has a larger value of phase transition temperature. We can also see from the right plot of free energy curves that the one with a first order phase transition get lower value of free energy than the one with second order phase transition.

Although we failed to find stable p+ip solutions, we still wish to find some qualitative estimation of the stability relation between the two solutions. One useful signal of stability can be taken as the critical back reaction strength $b_c$, beyond which the phase transition is first order. From the condensate and grand potential curves in Figure.~\ref{CondensateFreeE} and in Ref.~\cite{Nie:2016pjt}, we can see that the phase transition of the more stable p-wave solution change from second order to first order at a lower critical value of back reaction strength. We denote this critical value of back reaction strength for the p-wave and p+ip solutions as $b_{c-p}$ and $b_{c-pip}$ respectively. Between the two solutions sharing the same critical point, the more stable one always get a lower value of $b_c$. Therefore the stability relation of the two solutions can be concretely shown from the value of $b_{c-p}$ and $b_{c-pip}$.

In this paper, we focus on the influence of $m^2$ on the stability relation of the two solutions as well as phase structure. We show these results in the following subsections.
\subsection{$b_{c-p}$ v.s. $b_{c-pip}$}
In this subsection, we give the dependence of $b_{c-p}$ and $b_{c-pip}$ on the value of $m^2$. This will show a qualitative stability relation between the two solutions, and help to confirm that the p-wave solution is always more stable.

\begin{figure}\center
\includegraphics[height=4cm] {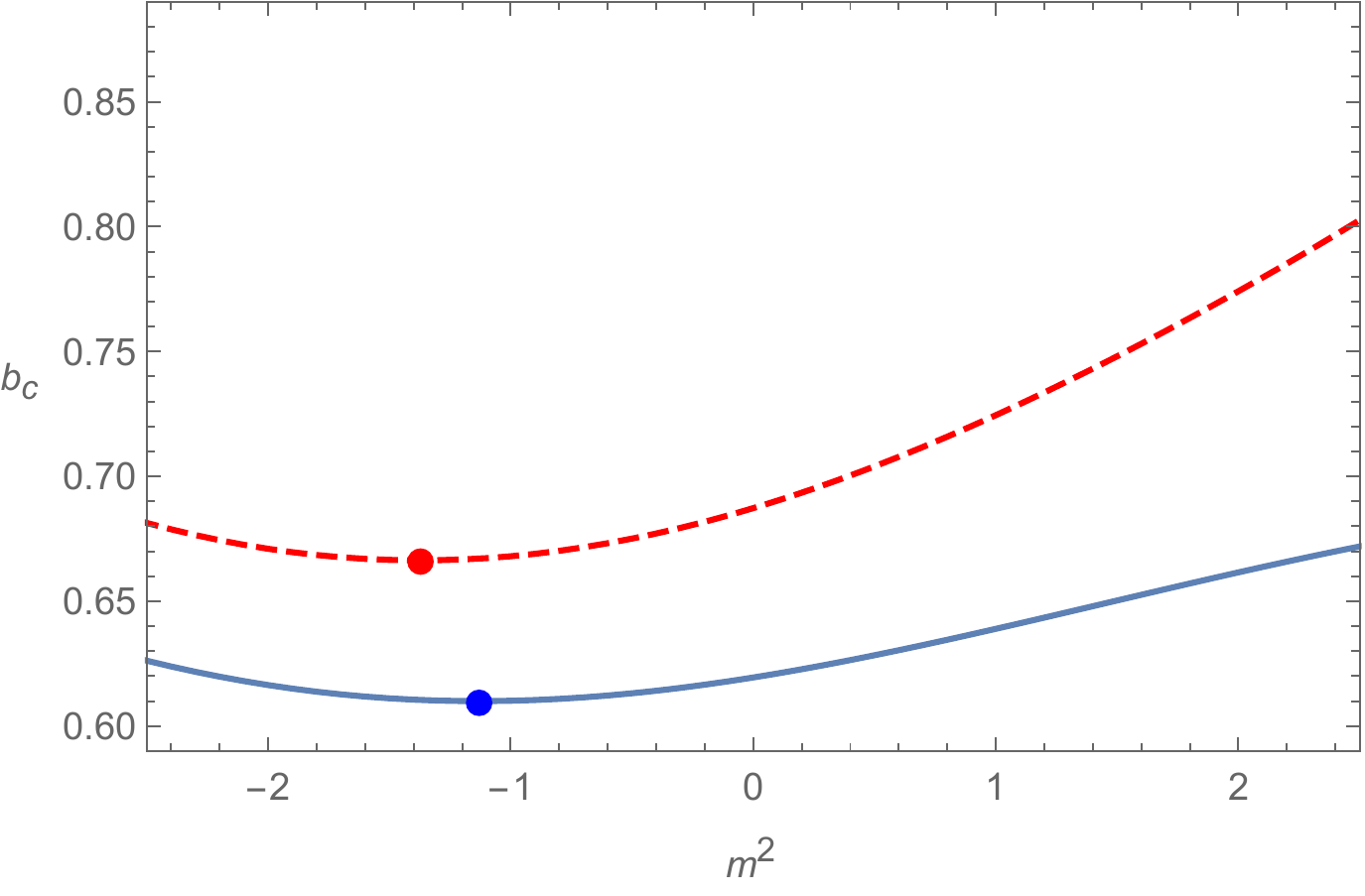}   \quad \quad
\includegraphics[height=4cm] {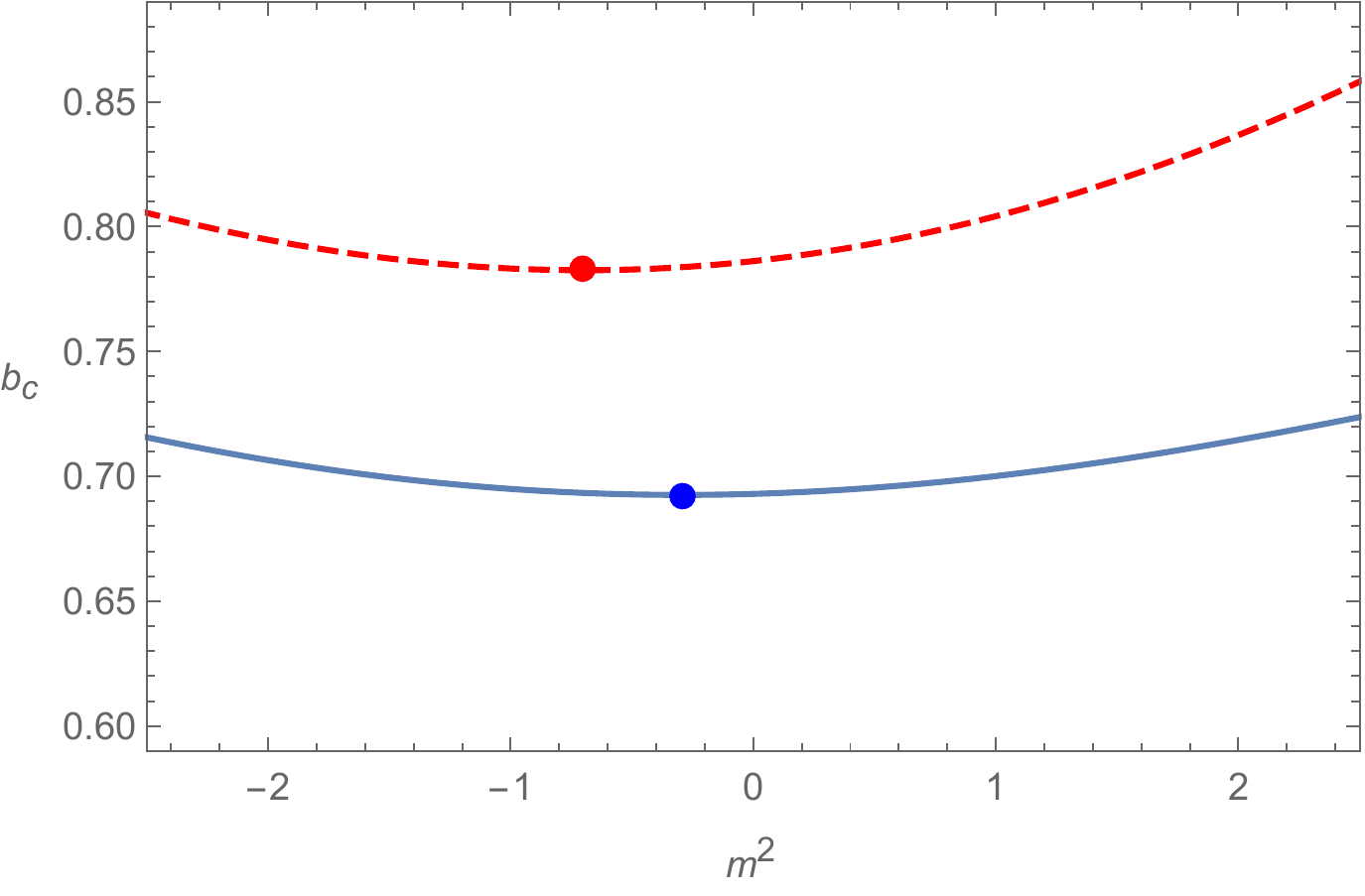}  
\caption{\label{bc-mg2}$b_c-m^2$ relations for $m_p^2=0$(Left plot) and $m_p^2=-3/16$(Right plot). The solid blue line denote the $b_c-m^2$ relation for the p-wave solution and the dashed red line denote the relation for p+ip solution, with the points on each line indicating the location of the minimum.
}
\end{figure}

We have set $\mu=3$ and $L=1, c_0=c_1=-2 c_2=1$, and consider two typical values $0$ and $-3/16$ for $m_p^2$ as in Ref.~\cite{Nie:2016pjt}. To show the influence of graviton mass on the stability relation, we varying the value of the graviton mass parameter $m^2$ and draw the two curves of $b_{c-p}$ and $b_{c-pip}$ in Figure.~\ref{bc-mg2}, where the left plot show the case of $m_p^2=0$ and the right plot show the one with $m_p^2=-3/16$. The solid blue line denote the $b_c-m^2$ relation for the p-wave solution and the dashed red line denote the relation for p+ip solution. The points on each line indicate the location of the minimum.

\begin{figure}\center
\includegraphics[height=4cm] {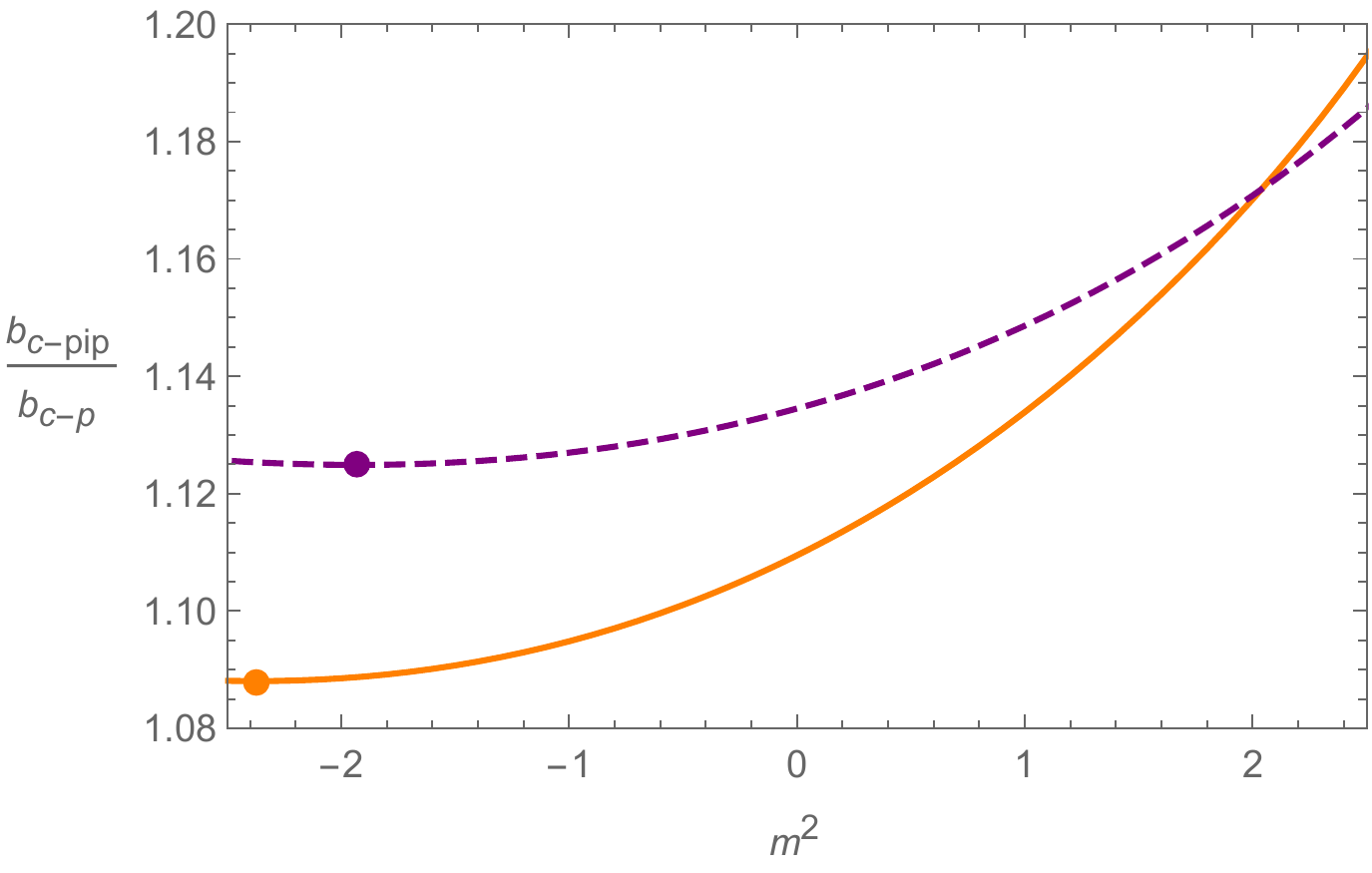}
\caption{\label{bcRation-mg2}$b_{c-pip}/b_{c-p}-m^2$ relations. The solid orange line is for the case with $m_p^2=0$ and the dashed purple line for the case with $m_p^2=-3/16$, with the two points on each line indicating the minimum. 
}
\end{figure}

At the beginning, we only consider positive values of $m^2$. We can see that in both the two cases ($m_p^2=0$ and $m_p^2=-3/16$), the value of $b_c$ for the two solutions all increase monotonically when the value of $m^2$ is increasing. To study the trend of the stability relation of the two solutions, we further draw the ratio $b_{c-pip}/b_{c-p}$ versus $m^2$ curves in Figure.~\ref{bcRation-mg2}.  The solid orange line is for the case with $m_p^2=0$ and the dashed purple line for the case with $m_p^2=-3/16$, with the two points on each line indicating the minimum. 

We can see that for positive values of $m^2$, the ratio $b_{c-pip}/b_{c-p}$ is larger than 1 and is monotonically increasing function of $m^2$. Therefore if we further decrease the value of $m^2$ to some negative value, it is possible that $b_{c-pip}/b_{c-p}$ becomes less than 1, which is a signal of a stable p+ip solution. In order to exclude this possibility, we extend our results to include negative values of $m^2$ and complete the left part of the curves in Figure.~\ref{bc-mg2} and Figure.~\ref{bcRation-mg2}. We can see that for both the two cases ($m_p^2=0$ and $m_p^2-3/16$), $b_{c-p}$, $b_{c-pip}$ and the ratio $b_{c-pip}/b_{c-p}$ all get a minimum at some negative value of $m^2$. Especially, the minimum of the ratio $b_{c-pip}/b_{c-p}$ is still larger than 1, indicating that it is not likely to make the p+ip solution win the competition against the p-wave one by tuning $m^2$.

The validity of negative value of $m^2$ can be understood from the following two aspects. On one side, the mass of the fields in AdS can get negative value above the B-F bound, such a bound may also be available for the graviton mass. On the other side, the minus sign of $m^2$ can be equivalently moved to the parameters $c_1$ and $c_2$. Thus the same results can be get from effectively considering positive value of $m^2$ and $c_1=-2 c_2=-1$.
\subsection{Critical temperature}
Because the p-wave solution always win the competition, the phase structure of this system is rather simple. It only include the normal phase in high temperature region and the p-wave phase in low temperature region. Another feature is that the phase transition becomes first order when the back reaction is strong enough. When the phase transition becomes first order, the phase transition point get a higher temperature than the ``critical point'' where the condensate emerge from the norm phase.

\begin{figure}\center
\includegraphics[height=4cm] {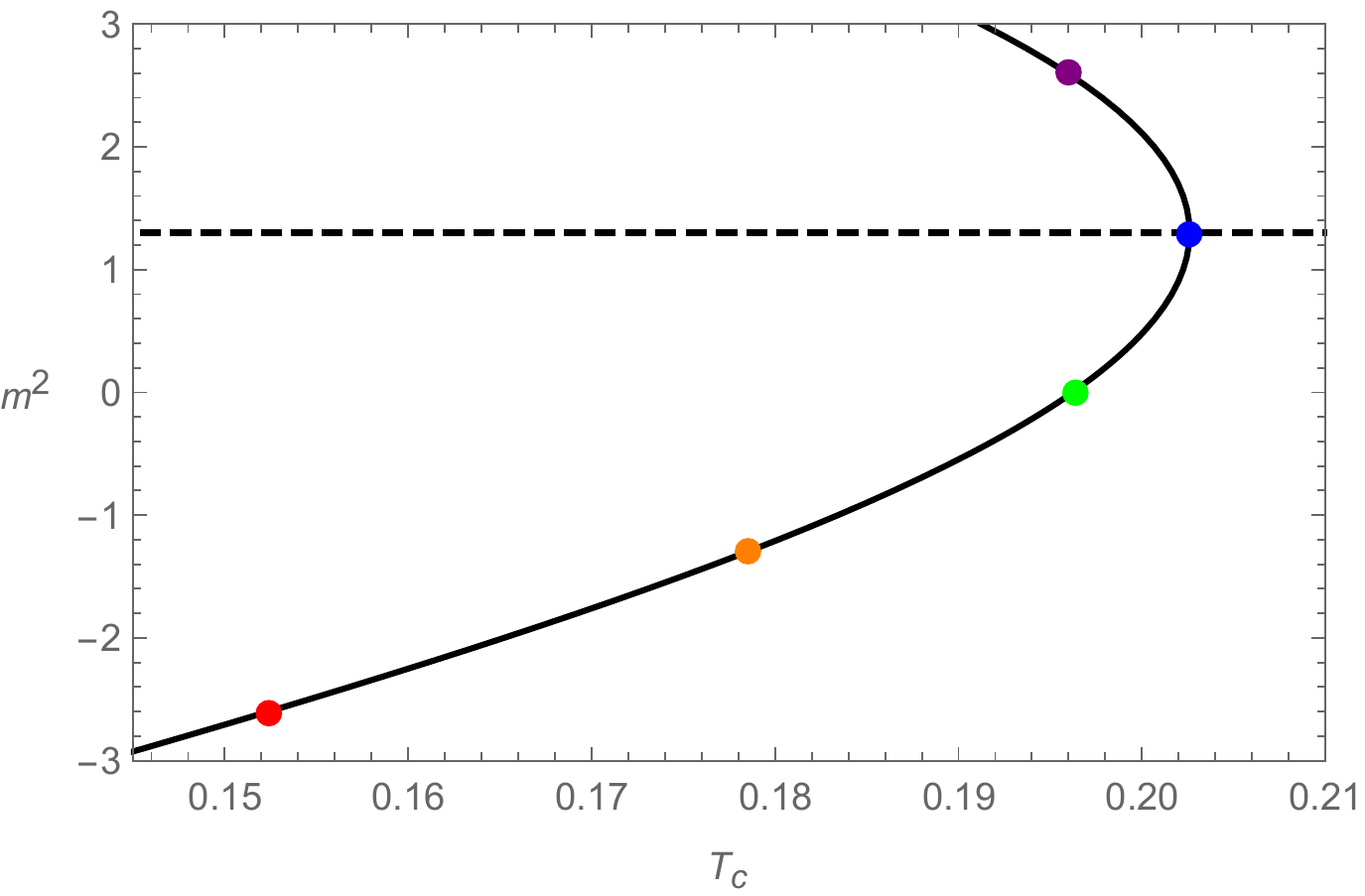}  \quad \quad
\includegraphics[height=4cm] {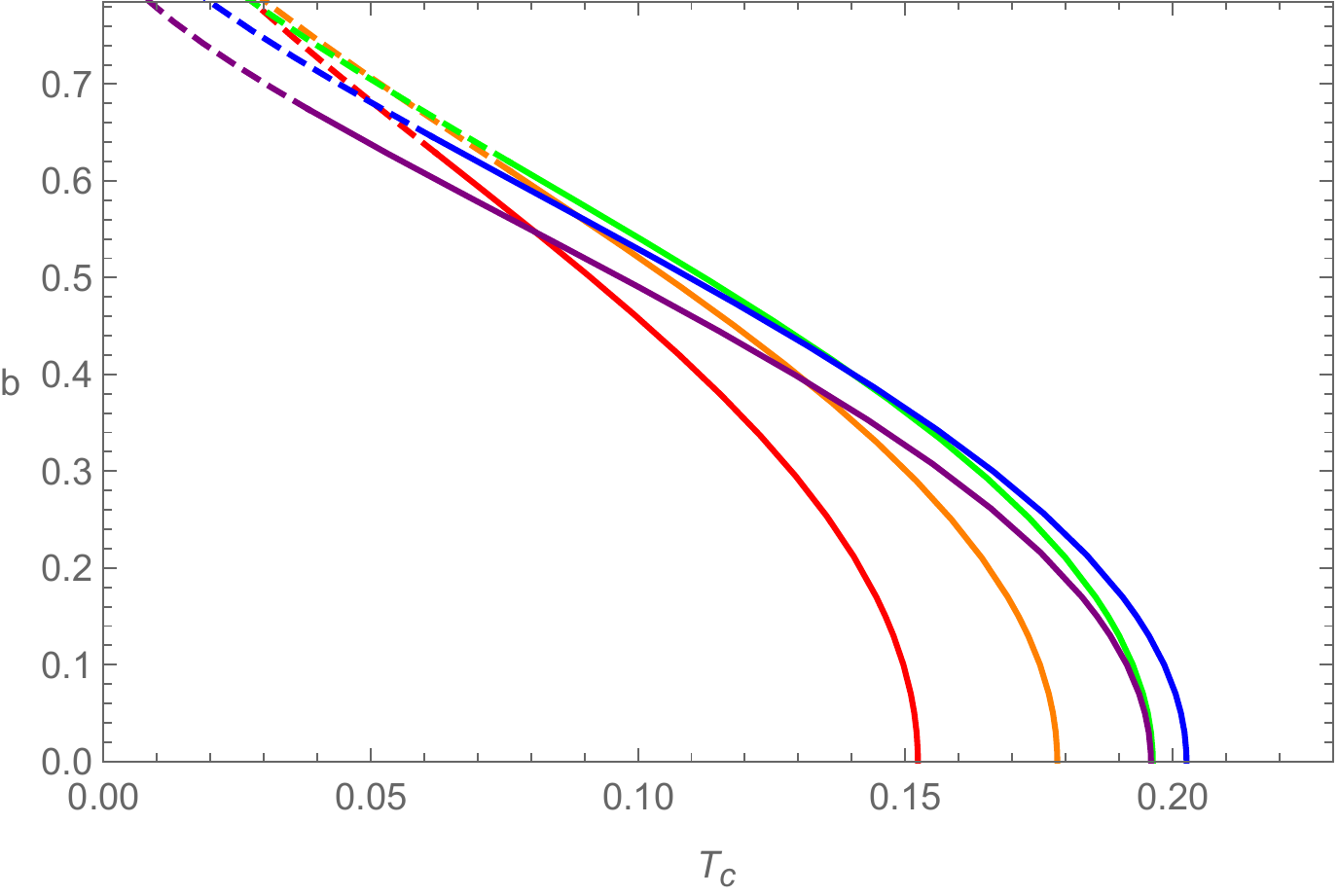}
\caption{\label{bTcPD} $m^2-T_c$ relation at probe limit(Left plot) and $b-T_c$ relations(Right plot) in the case $m_p^2=0$. In the left plot, we show the $T_c-m^2$ relation of the p-wave phase when back reaction can be neglected, while in the right plot we show the $b-T_c$ curves for the critical points with 5 values of $m^2$. The dashed black horizontal line in the left figure indicate the maximum critical temperature at $m^2=1.30$. In both the two plots, the five colors \{red, orange, green, blue, purple\} are used to denote five values of $m^2$: \{-2.6, -1.3, 0.01, 1.3, 2.6\}.
}
\end{figure}

We studied the impact of graviton mass parameter $m^2$ on the critical temperature of p-wave condensate only for $m_p^2=0$, because the other case $m_p^2=-3/16$ involve 0th order phase transitions at lower temperature, which make it complicated to get the phase diagram and is not the focus of this work. Because the back reaction strength also affect the critical temperature, we start from the probe limit $b=0$ and draw the relation of $m^2-T_c$ in the left plot of Figure.~\ref{bTcPD}. This plot is also a 2D phase diagram in probe limit.

We can see from this plot that the critical temperature get a maximum at a positive value of $m^2=1.30$. To under stand this non-monotonic behavior, we can see the formula Eq.~(\ref{Temperature}) for temperature. In probe limit, only the first term proportional to $r_h$ and two terms proportional to $m^2$ left in that formula. We confirmed that when $m^2$ is increasing, the value of $r_h$ and the first term in Eq.~(\ref{Temperature}) for the critical point decrease monotonically. However, the increasing of $m^2$ has an effect of increasing the critical temperature through the two terms in Eq.~(\ref{Temperature}). As a result, the final dependence of $T_c$ on $m^2$ show a non-monotonic behavior combining the above two effects.

To get more information away from the probe limit, we choose five values of $m^2$ and show the $b-T_c$ curves in the right plot of Figure.~\ref{bTcPD}. We use \{Red, Orange, Green, Blue, Purple\} to denote the lines with $m^2=$ \{-2.6, -1.3, 0.01, 1.3, 2.6\} respectively. We also mark the five points with the selected value of $m^2$ in the left plot with the same color assignment. In the right plot of Figure.~\ref{bTcPD}, the solid lines are all real boundary of the p-wave phase, therefore the five solid lines also describe 2D slices of the phase diagram at different values of $m^2$. The dashed lines denote the ``critical point'' of the first order phase transition, and has a temperature lower than the real phase transition point.

Because of the non-monotonic effect of $m^2$, the relation of the five colored lines are complicated. In general, we can see that all these lines show a decreasing of critical temperature when the back reaction strength is increasing. The one with a larger value of $m^2$ has more decreasing of critical temperature, which can be attributed to the last two terms in Eq.~(\ref{Temperature}).
\section{\bf Conclusions and discussions}\label{sect:conclusion}
In this paper, we studied the complex vector p-wave mode within dRGT massive gravity. We considered the full back reaction and study the competition between the p-wave and p+ip solutions, and find that the p-wave solution still always win the competition. We also compare the value of critical back reaction strength, beyond which the phase transition become first order, to show a qualitative stability relation. We also give the value of critical temperature at different values of graviton mass parameter $m^2$ and back reaction strength $b$.

In the case of dRGT massive gravity, a key scaling symmetry involve the parameter $c_0$, therefore one can not use the scaling trick to easily get dependence of temperature with a fixed value of $c_0$. To solve this problem, we take varying value of $r_h$ in our numerical work to get the varying temperature directly, which is a more general numerical treatment.

Since the p+ip solution still failed to win the competition against the p-wave one, we can continue exploring this competition in new setups. With in this study, we find that the ratio $b_{c-pip}/b_{c-p}$ can be a convenient signal to be used in future study.
\begin{acknowledgments}
ZYN would like to thank Qi-Yuan Pan for useful discussions and suggestions.
This work was supported in part by the National Natural Science Foundation of China under Grant Nos. 11565017, 11965013, and 11575083. 
ZYN is supported in part by Yunnan Ten Thousand Talents Plan Young \& Elite Talents Project.
\end{acknowledgments}



\end{document}